\documentclass[conference,a4paper]{IEEEtran}

\usepackage[T1]{fontenc}
\usepackage{amsmath,amssymb}
\usepackage{booktabs}
\usepackage{array}
\usepackage{multirow}
\usepackage{graphicx}
\usepackage{xcolor}
\usepackage{microtype}
\usepackage{balance}
\usepackage{xspace}
\usepackage{tikz}
\usetikzlibrary{positioning,arrows.meta,calc}
\usepackage[hidelinks]{hyperref}
\usepackage{url}

\newcommand{\hiqe}{\textsc{HiQE}\xspace}
\newcommand{\vect}[1]{\mathbf{#1}}

\newcommand{\NFCorpusBestRelative}{4.81\%}

\newcommand{\TrecCovidBestRelative}{8.92\%}

\newcommand{\SciDocsBestRelative}{9.47\%}

\title{Generated Query Expansion Still Helps Strong Sparse Retrieval: A Controlled Study with SPLADE-v3}

\author{\IEEEauthorblockN{
Ryan C. Barron\IEEEauthorrefmark{1},
Cade W. Trotter\IEEEauthorrefmark{2},
Maksim E. Eren\IEEEauthorrefmark{1},
Kim \O. Rasmussen\IEEEauthorrefmark{3},
Liz D. Miller\IEEEauthorrefmark{4}
Benjamin J. Migliori\IEEEauthorrefmark{5}
}
\IEEEauthorblockA{
\IEEEauthorrefmark{1}Computational Intelligence \& Modeling, Los Alamos National Laboratory, Los Alamos, New Mexico, USA. \\
\IEEEauthorrefmark{2}Modeling and Observations of Earth Systems, Los Alamos National Laboratory, Los Alamos, New Mexico, USA. \\
\IEEEauthorrefmark{3}Fluid Dynamics and Solid Mechanics, Los Alamos National Laboratory, Los Alamos, New Mexico, USA. \\
\IEEEauthorrefmark{4}Intelligence \& Systems Analysis, Los Alamos National Laboratory, Los Alamos, New Mexico, USA. \\
\IEEEauthorrefmark{5}Advanced Research in Cyber Systems, Los Alamos National Laboratory, Los Alamos, New Mexico, USA.
}

\thanks{U.S. Government work not protected by U.S. copyright.}
}

\begin{document}
\maketitle

\begin{abstract}
Scientific queries are often brief, while relevant papers use specialized vocabulary. Generated query expansion can bridge this mismatch, but earlier work suggests that its value shrinks as the underlying retriever becomes stronger. We test the four generated formats of term lists, a pseudo-document, multiple pseudo-references, and corpus-steered text all together with SPLADE-v3 on NFCorpus, TREC-COVID, and SciDocs. Every condition searches the same frozen document index and follows the same query-side integration rule and 256-dimension budget, isolating the effect of the added content. All twelve method-collection comparisons improve aggregate nDCG@10, with best relative gains of 4.81\%, 8.92\%, and 9.47\%. Eleven remain significant after Holm correction. The gain persists in 103 of 114 interpolation settings, including every setting that assigns at least 30\% of the mixture weight to the original query. Shuffled-text and non-contextual lexical-bag controls also remain above baseline in all 24 aggregate comparisons, showing that the added vocabulary carries most of the benefit. A corpus-induced typed concept graph, by contrast, produces no consistent gain, and its relation, depth, validation, random, and gating controls do not rescue it. Generated vocabulary can therefore complement a strong learned sparse retriever, provided that the original query remains strongly represented.
\end{abstract}


\section{Introduction}
Scientific search has a persistent vocabulary problem.
A user may describe a phenomenon, material, or desired outcome in a few ordinary words, while the relevant literature uses specialized entities, method names, and domain terminology.
Query expansion tries to close this gap by adding words that the user did not supply.
Classical methods derive those words from judged or initially retrieved documents \cite{rocchio1971relevance,lavrenko2001relevance,abduljaleel2004umass}. Newer methods ask language models to generate keywords, hypothetical documents, or several pseudo-references \cite{jagerman2023query,wang-etal-2023-query2doc,gao2023hyde,zhang-etal-2024-exploring-best}.
The opportunity is clear, but so is the risk: generated text may introduce useful terminology, or it may pull the ranking away from the user's intent.

This tradeoff is especially sharp for learned sparse retrieval.
SPLADE already maps queries and documents to weighted vocabulary terms, including terms inferred from context, while retaining efficient inverted-index search \cite{formal2021splade,formal2021spladev2}.
SPLADE-v3 is a strong checkpoint in this family \cite{lassance2024spladev3}.
If its query representation already expands the user's wording, what can an external generator still add?
Prior work makes the answer uncertain: across many settings, generative expansion helps weaker retrievers more and can harm the strongest ones \cite{weller-etal-2024-generative}.

We answer this question with a controlled comparison.
On NFCorpus, TREC-COVID, and SciDocs, we add four forms of generated content to the query side of SPLADE-v3: a short term list, one pseudo-document, four pseudo-references, and corpus-steered text.
Every condition searches the same frozen SPLADE-v3 document index, uses the same candidate depth, and obeys the same sparse query budget.
We also test \hiqe, a structured alternative that selects concepts induced from the corpus, traverses typed relations, and projects the selected concepts into the same SPLADE vocabulary space.
This comparison helps distinguish a benefit specific to generated vocabulary from a generic benefit of adding related terms.

The answer is consistent across the three collections.
Every generated format improves aggregate nDCG@10, with best relative gains of \NFCorpusBestRelative\ on NFCorpus, \TrecCovidBestRelative\ on TREC-COVID, and \SciDocsBestRelative\ on SciDocs.
The gains persist across a broad range of mixture weights and survive controls that shuffle the generated text or reduce it to a non-contextual lexical bag.
The concept graph, however, does not show a comparable pattern.
Together, the controls point to a straightforward explanation: the generator contributes useful scientific vocabulary, but that vocabulary works best as a supplement to a strongly preserved original query.

The paper makes three contributions.
\begin{enumerate}
  \item \textbf{A controlled strong-retriever test.} Four generated formats improve one frozen SPLADE-v3 document index across three scientific-search collections.
  \item \textbf{Evidence about why the gains persist.} Weight sweeps, paired tests, and same-content controls show that the benefit is broad and depends more on added vocabulary than on exact generated word order.
  \item \textbf{A structured counterpoint.} A corpus-induced concept graph and its relation, depth, random, and gating controls fail to reproduce the generated gains, showing that query-side term addition alone is not sufficient.
\end{enumerate}

\section{Related Work}

\subsection{Generated Query Expansion}

Classical query expansion derives new vocabulary from the collection.
Rocchio-style feedback moves the query toward judged or pseudo-relevant documents \cite{rocchio1971relevance}. Relevance models estimate terms from initially retrieved evidence \cite{lavrenko2001relevance,abduljaleel2004umass}.
These methods are corpus-grounded, but they can amplify first-stage errors when the feedback set mixes different interpretations of the query.

Language models provide another source of vocabulary.
They can generate keyword lists \cite{jagerman2023query}, a single hypothetical passage as in Query2doc \cite{wang-etal-2023-query2doc}, a document representation as in HyDE \cite{gao2023hyde}, or multiple pseudo-references as in MuGI-style prompting \cite{zhang-etal-2024-exploring-best}.
Corpus-steered methods condition generation on retrieved evidence \cite{lei-etal-2024-corpus}.
Each format covers the information need differently: lists are compact, passages can connect concepts, multiple references can represent several facets, and corpus steering can constrain the vocabulary while inheriting first-stage bias.

These benefits are not guaranteed.
Weller et al. find that generative expansion generally helps weaker retrieval systems more than stronger ones and often harms the strongest retrievers \cite{weller-etal-2024-generative}.
A generator may also reproduce benchmark-specific evidence encountered during training instead of providing a transferable reformulation \cite{yoon-etal-2025-hypothetical}.
We hold the generated texts fixed so that our retrieval experiments measure integration effects rather than generation variability.

\subsection{Learned Sparse Retrieval and Integration}

SPLADE learns contextual lexical expansion while retaining sparse retrieval \cite{formal2021splade,formal2021spladev2}, where SPLADE-v3 is a strong modern baseline in this family \cite{lassance2024spladev3}.
Because its query and document vectors already contain learned expansion, an external reformulator must contribute genuinely complementary evidence.
We isolate that contribution by freezing the document side and changing only the query representation.

Original and expanded evidence can be combined in several ways.
Reciprocal rank fusion merges ranked lists without requiring a common score scale \cite{cormack2009reciprocal}, and Exp4Fuse applies route-level fusion to language-model expansion for sparse retrieval \cite{liu-zhang-2025-exp4fuse}.
QuDAR instead assigns adaptive weights across original and expanded queries and across sparse and dense retrieval \cite{kim-etal-2026-qudar}.
Our study focuses on fixed query-side interpolation because the shared sparse space supports a tightly controlled comparison between generated text and corpus-induced concepts.

\subsection{Structured Expansion and Evaluation}

Knowledge-aware expansion and scientific taxonomy construction motivate explicit concept structure \cite{xia-etal-2025-knowledge,kargupta-etal-2025-taxoadapt}.
Typed edges can represent relations among candidate additions, but a graph also adds several possible failure points: phrase extraction, relation induction, query-to-concept mapping, traversal, and projection.
Even a well-formed graph may contain edges that are irrelevant to ranking.
We therefore compare typed traversal with flat, relation, depth, validation, and random controls.

Finally, aggregate means can hide gains concentrated in a few topics.
Following established information-retrieval practice \cite{smucker-2007-comparison}, we supplement aggregate effectiveness with paired intervals, corrected p-values, standardized effects, and wins/ties/losses.

\section{Study Design and Methods}
\label{sec:design}
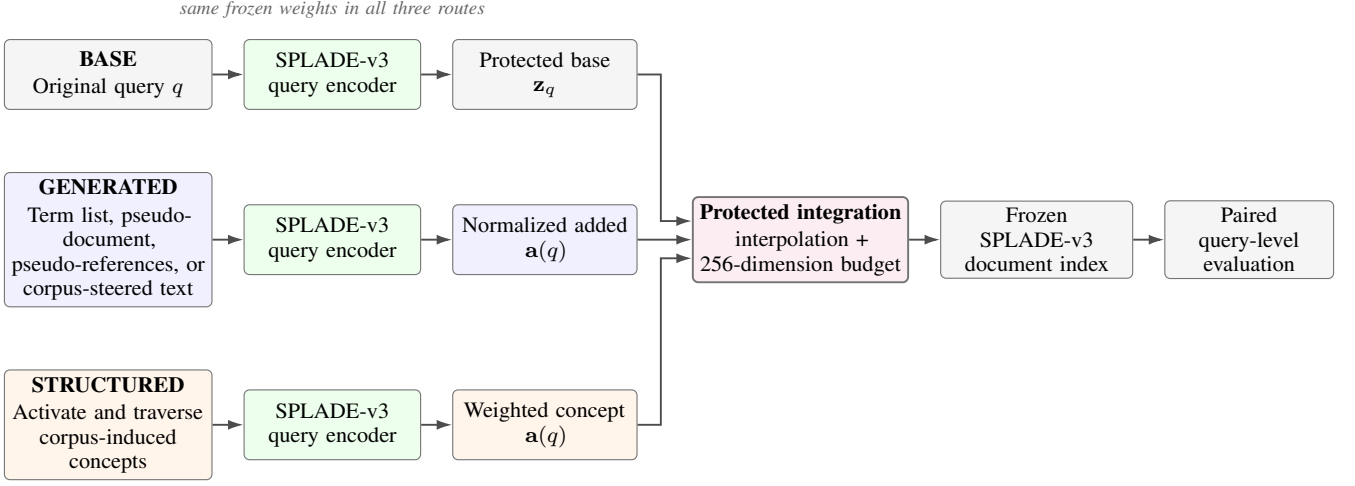
\begin{figure*}[t]
\centering

\resizebox{0.97\textwidth}{!}{%
\begin{tikzpicture}[
    node distance=5mm and 4mm,
    box/.style={draw=black!55, rounded corners=2pt, align=center, minimum height=9mm, inner sep=2.5pt, font=\footnotesize},
    input/.style={box, text width=25mm},
    encoder/.style={box, text width=21mm, fill=green!7},
    vectorbox/.style={box, text width=22mm},
    integration/.style={box, text width=26mm, fill=purple!7, line width=0.7pt},
    downstream/.style={box, text width=23mm, fill=gray!8},
    evalbox/.style={box, text width=20mm, fill=gray!8},
    arrow/.style={-{Latex[length=2mm,width=1.3mm]}, line width=0.7pt, draw=black!70}
]

\node[
    input,
    fill=gray!8
] (q) {
    \textbf{BASE}\\[1pt]
    Original query $q$
};

\node[
    encoder,
    right=of q
] (baseenc) {
    SPLADE-v3\\
    query encoder
};

\node[
    vectorbox,
    fill=gray!8,
    right=of baseenc
] (basevec) {
    Protected base\\
    $\mathbf{z}_q$
};

\node[
    input,
    fill=blue!6,
    below=8mm of q
] (gen) {
    \textbf{GENERATED}\\[1pt]
    Term list, pseudo-document,\\
    pseudo-references, or\\
    corpus-steered text
};

\node[
    encoder,
    right=of gen
] (genenc) {
    SPLADE-v3\\
    query encoder
};

\node[
    vectorbox,
    fill=blue!6,
    right=of genenc
] (genvec) {
    Normalized added\\
    $\mathbf{a}(q)$
};

\node[
    input,
    fill=orange!8,
    below=8mm of gen
] (graph) {
    \textbf{STRUCTURED}\\[1pt]
    Activate and traverse\\
    corpus-induced concepts
};

\node[
    encoder,
    right=of graph
] (graphenc) {
    SPLADE-v3\\
    query encoder
};

\node[
    vectorbox,
    fill=orange!8,
    right=of graphenc
] (graphvec) {
    Weighted concept\\
    $\mathbf{a}(q)$
};

\node[
    integration,
    right=7mm of genvec
] (merge) {
    \textbf{Protected integration}\\[1pt]
    interpolation +\\
    256-dimension budget
};

\node[
    downstream,
    right=of merge
] (index) {
    Frozen SPLADE-v3\\
    document index
};

\node[
    evalbox,
    right=of index
] (eval) {
    Paired query-level\\
    evaluation
};

\draw[arrow] (q) -- (baseenc);
\draw[arrow] (baseenc) -- (basevec);

\draw[arrow] (gen) -- (genenc);
\draw[arrow] (genenc) -- (genvec);

\draw[arrow] (graph) -- (graphenc);
\draw[arrow] (graphenc) -- (graphvec);

\draw[arrow]
    (basevec.east)
    -- ++(3mm,0)
    |- ([yshift=2.4mm]merge.west);

\draw[arrow]
    (genvec.east)
    -- (merge.west);

\draw[arrow]
    (graphvec.east)
    -- ++(3mm,0)
    |- ([yshift=-2.4mm]merge.west);

\draw[arrow] (merge) -- (index);
\draw[arrow] (index) -- (eval);

\node[
    font=\scriptsize\itshape,
    text=black!60,
    above=1.5mm of baseenc
] {
    same frozen weights in all three routes
};

\end{tikzpicture}%
}

\caption{Controlled study design. The original query, generated expansion, and corpus-induced concept expansion are encoded using the same frozen SPLADE-v3 query model. Added evidence is combined with the protected original-query representation under a fixed 256-dimension sparse budget, and every condition searches the same frozen document index before paired query-level evaluation.}
\label{fig:study-design}

\end{figure*}

The experiments follow three research questions:
\begin{description}
  \item[\textbf{RQ1}] Does generated expansion improve a frozen SPLADE-v3 retriever?
  \item[\textbf{RQ2}] Does any gain survive changes in mixture weight and in the representation of the same generated content?
  \item[\textbf{RQ3}] Can corpus-induced concepts, typed relations, or selective gates produce a comparable gain?
\end{description}
Figure~\ref{fig:study-design} summarizes the shared retrieval pipeline.

\subsection{Shared Frozen-Index Pipeline}
Let $q$ be a query and $d$ a document.
A frozen SPLADE-v3 encoder produces the base query vector $\vect{z}_q$ and document vectors $\vect{z}_d$.
We build the document vectors once and reuse them in every matched condition.
Only the query vector changes, and every condition scores documents as $s(d,q)=\widetilde{\vect{z}}_q^{\top}\vect{z}_d$.
For generated method $m$, let $\bar{\vect{z}}_q$ and $\bar{\vect{a}}_m(q)$ denote row-wise $\ell_1$-normalized base and added SPLADE vectors.
We combine them as
\begin{equation}
 \widetilde{\vect{z}}_q = \operatorname{Budget}_{256}
 \left(\alpha_m\bar{\vect{z}}_q + (1-\alpha_m)\bar{\vect{a}}_m(q)\right),
 \label{eq:integration}
\end{equation}
Here, $\alpha_m$ controls the balance: larger values preserve more of the original query, while smaller values give the expansion more influence.
The final vector contains at most 256 active dimensions.
We protect the original query within that budget: if the base vector has fewer than 256 dimensions, only addition-only dimensions are pruned. If the budget is exceeded, we retain its 256 highest-weight dimensions.
Generated content and graph concepts are encoded separately, so they do not consume the original query's transformer input length.
Thus, the document representation, scorer, candidate depth, query budget, and metric implementation remain fixed across the matched conditions.

The configured original-query weights are $\alpha_m=0.65$ for Flat LLM-QE and CSQE and $0.50$ for Query2doc and MuGI-style expansion.
Section~\ref{sec:weight-robustness} tests sensitivity to this choice.
Graph projection uses $\lambda_{\mathrm{proj}}=0.75$ under the same 256-dimension budget.

\subsection{Generated Expansion Conditions}
The four generated conditions differ only in the form of the added text.
Flat LLM-QE produces up to twelve search terms.
Query2doc produces one hypothetical relevant passage \cite{wang-etal-2023-query2doc}.
MuGI-style expansion produces four pseudo-references, providing several views of the information need \cite{zhang-etal-2024-exploring-best}.
Corpus-Steered Query Expansion (CSQE) first retrieves evidence, then generates pivotal sentences and knowledge terms conditioned on that evidence \cite{lei-etal-2024-corpus}.
In every case, SPLADE encodes the added text and Eq.~\ref{eq:integration} combines it with the protected base query.

The run archive preserves the generated texts and the LLM response cache.
The cache records the served model identifier \texttt{gpt-oss-120b}, the system and user prompts, request parameters (including temperature 0.0 and the output limit), and raw responses.
The retrieval experiments can therefore be rerun from the exact fixed expansions.
Independent regeneration should also use the provider and model-revision metadata distributed with the release.
The four generated conditions are not matched for model calls, token use, or latency. The study compares retrieval effectiveness rather than generation efficiency.

\subsection{Interpolation and Same-Content Controls}
We sweep $\alpha\in\{0.1,0.2,\ldots,0.9\}$ and also evaluate the configured $0.65$ for Flat LLM-QE and CSQE.
At the configured $\alpha_m$, two controls preserve the generated content while changing its representation.
The first deterministically shuffles whitespace tokens before applying the same SPLADE encoder, then the second forms an $\ell_1$-normalized term-frequency bag from the same tokenizer wordpieces, removing contextual SPLADE expansion.
MuGI pseudo-references remain separately normalized and averaged.
Alternative representations use the same paired tests, with Holm correction across eight comparisons per collection.

\subsection{Corpus-Induced Structured Comparison}
The structured condition extracts one- to three-token TF--IDF keyphrases from scientific titles and abstracts, then records document--concept assignments.
Instead of generating text, \hiqe expands a query with related concepts induced from the corpus.
Typed broader, sibling, and related edges are derived from phrase relations, dense similarity, shared parents, co-occurrence, and corpus support.
For the publication run, edge validation is deterministic and grounded in corpus evidence. A separate experimental variant adds LLM-based evidence judgments, but those judgments are not used in the main results.

The query mapper combines dense similarity, lexical overlap, and aliases.
It retains at most four seed concepts, traverses at most two edges, and keeps at most twelve expansion concepts.
For a path $\pi=(e_1,\ldots,e_\ell)$ from seed $c_0$ to concept $c$, the path score is
\begin{equation}
 \Omega(\pi\mid q)=p(c_0\mid q)
 \left(\prod_{j=1}^{\ell}\alpha_{r(e_j)}v(e_j)\right)
 \kappa^{\max(0,\ell-1)},
\end{equation}
where $v(e)$ is edge confidence, $\alpha_r$ is a relation prior, and the depth decay is $\kappa=0.75$.
Before frozen-SPLADE encoding, the projection text combines the concept name with optional aliases and snippets from representative documents.
The resulting graph vector is mixed with the base query using $\lambda_{\mathrm{proj}}=0.75$ under the 256-dimension budget.
Controls remove validation, equalize relation weights, restrict relation families, reduce traversal depth, flatten concept scores, or substitute random count-matched concepts.

A heuristic gate uses hand-set activation and confidence criteria.
A class-balanced logistic-regression gate is trained on two collections and applied to the held-out third collection in a leave-one-dataset-out protocol.
We report gate activation, harmful ungated expansions avoided, beneficial ungated expansions missed, and the gap to a per-query oracle that selects the better of unexpanded SPLADE-v3 and ungated graph expansion.
The oracle measures the maximum gain obtainable from perfect per-query selection of the two already-computed rankings.

\subsection{Collections, Baselines, and Statistical Analysis}
We evaluate three scientific-search collections distributed through the BEIR benchmark suite: NFCorpus, TREC-COVID, and SciDocs \cite{thakur2021beir,boteva2016full,voorhees2021trec}.
The three collections contain 3,633, 171,332, and 25,657 documents and 323, 50, and 1,000 test queries, respectively.
We report each collection separately because their query counts and relevance structures differ substantially.

The main comparison includes BM25 \cite{robertson2009probabilistic}, RM3, TAS-B dense retrieval, BM25+dense reciprocal-rank fusion, ColBERTv2 late interaction \cite{santhanam-etal-2022-colbertv2}, SPLADE-v3, the four generated conditions, and three graph conditions.
BM25 provides a lexical baseline and RM3 adds pseudo-relevance feedback.
TAS-B is a dense bi-encoder, reciprocal-rank fusion combines the BM25 and dense lists, and ColBERTv2 is a late-interaction retriever.
SPLADE-v3 is the frozen learned-sparse baseline directly matched to our query-expansion conditions.
Candidate depth is 1,000.
nDCG@10 is the primary metric. MRR@10, MAP@10, Precision@10, Recall@10, Recall@100, and candidate recall are secondary aggregate measures.
nDCG@10 measures the quality of the top-10 ranking. MRR@10 emphasizes how early the first relevant result appears. MAP@10 summarizes precision across relevant results in the top 10. Recall@100 measures deeper coverage.

From aligned per-query nDCG@10 values, we report mean difference, relative change, Cohen's paired $d_z$, wins/ties/losses, a 95\% interval from 20,000 paired bootstrap resamples, and a two-sided test from 100,000 paired sign randomizations (seed 13).
Generated-method p-values use Holm correction across the four methods within each collection. The three ungated graph comparisons use Bonferroni correction across collections.
A win or loss is defined by the sign of the per-query nDCG@10 difference, while exact equality is retained as a tie.
No manual semantic annotation of graph edges was conducted.

\section{Results}
Each subsection answers one research question first, then presents the evidence supporting that answer.

\subsection{RQ1: Does Generated Expansion Improve SPLADE-v3?}
\noindent\textbf{Answer to RQ1.}
Yes. All four generated formats improve the matched SPLADE-v3 baseline on all three collections. The best relative nDCG@10 gains are 4.81\% on NFCorpus, 8.92\% on TREC-COVID, and 9.47\% on SciDocs.
\begin{table*}[t]
\centering
\caption{First-stage effectiveness, with nDCG@10 as the primary metric, R@100 as deeper coverage, and bold marking the best value in each dataset column.}
\label{tab:main-results}
\footnotesize
\setlength{\tabcolsep}{3.5pt}
\begin{tabular}{lcccccc}
\toprule
& \multicolumn{2}{c}{NFCorpus} & \multicolumn{2}{c}{TREC-COVID} & \multicolumn{2}{c}{SciDocs} \\
\cmidrule(lr){2-3}\cmidrule(lr){4-5}\cmidrule(lr){6-7}
Method & nDCG@10 & R@100 & nDCG@10 & R@100 & nDCG@10 & R@100 \\
\midrule
BM25 & 0.3231 & 0.2457 & 0.5696 & 0.1091 & 0.1490 & 0.3477 \\
RM3 & 0.3465 & \textbf{0.3229} & 0.5635 & 0.1168 & 0.1491 & 0.3620 \\
TAS-B dense & 0.2755 & 0.2487 & 0.3087 & 0.0335 & 0.1406 & 0.3222 \\
BM25 + dense RRF & 0.3342 & 0.2819 & 0.5511 & 0.0894 & 0.1690 & 0.3848 \\
ColBERTv2 & 0.3392 & 0.2803 & 0.7100 & 0.1307 & 0.1482 & 0.3543 \\
\midrule
SPLADE-v3 & 0.3596 & 0.2971 & 0.7275 & 0.1388 & 0.1579 & 0.3709 \\
\midrule
Flat LLM-QE & 0.3715 & 0.3060 & 0.7850 & 0.1532 & 0.1688 & \textbf{0.3973} \\
Query2doc & 0.3724 & 0.3102 & \textbf{0.7924} & 0.1533 & 0.1673 & 0.3923 \\
MuGI-style & 0.3701 & 0.3130 & 0.7840 & \textbf{0.1533} & 0.1668 & 0.3890 \\
CSQE & \textbf{0.3769} & 0.3155 & 0.7556 & 0.1481 & \textbf{0.1729} & 0.3916 \\
\midrule
HiQE, ungated & 0.3609 & 0.2961 & 0.7149 & 0.1379 & 0.1585 & 0.3717 \\
HiQE, heuristic gate & 0.3596 & 0.2971 & 0.7275 & 0.1388 & 0.1579 & 0.3709 \\
HiQE, learned gate & 0.3603 & 0.2958 & 0.7149 & 0.1379 & 0.1579 & 0.3709 \\
\bottomrule
\end{tabular}
\end{table*}

Table~\ref{tab:main-results} shows that CSQE is strongest on NFCorpus and SciDocs, while Query2doc is strongest on TREC-COVID.
The corresponding nDCG@10 gains over unexpanded SPLADE-v3 are \NFCorpusBestRelative, \TrecCovidBestRelative, and \SciDocsBestRelative.
No single format wins everywhere: CSQE is the weakest generated condition on TREC-COVID, and Query2doc does not lead the other two collections.
The effect is therefore shared across generation styles rather than driven by one universally best method.

The broader baseline comparison is less uniform.
CSQE exceeds the strongest listed non-SPLADE nDCG@10 baseline by $0.0304$ on NFCorpus and $0.0039$ on SciDocs; Query2doc exceeds ColBERTv2 by $0.0823$ on TREC-COVID.
For Recall@100, however, RM3 remains best on NFCorpus ($0.3229$), while Flat LLM-QE is best on SciDocs ($0.3973$).
Generated expansion reliably improves the matched SPLADE-v3 run, but the best system still depends on the collection and metric.

\begin{table*}[t]
\centering
\caption{Relative improvement (\%) over SPLADE-v3 for the strongest generated nDCG@10 condition on each collection.}
\label{tab:cross-metric-gains}
\small
\setlength{\tabcolsep}{5.5pt}
\begin{tabular}{llrrrrrrr}
\toprule
Dataset & Method & nDCG@10 & MRR@10 & MAP@10 & P@10 & R@10 & R@100 & Cand. recall \\
\midrule
NFCorpus & CSQE & +4.81 & +5.10 & +5.30 & +4.45 & +1.81 & +6.20 & +6.28 \\
TREC-COVID & Query2doc & +8.92 & +3.74 & +6.48 & +3.91 & +6.99 & +10.44 & +5.11 \\
SciDocs & CSQE & +9.47 & +9.16 & +12.58 & +7.38 & +7.39 & +5.57 & +3.14 \\
\bottomrule
\end{tabular}
\end{table*}

\paragraph{Evidence across metrics.}
The strongest generated condition on each collection also improves all six secondary measures in Table~\ref{tab:cross-metric-gains}.
Notably, Recall@100 rises by 6.20\% on NFCorpus, 10.44\% on TREC-COVID, and 5.57\% on SciDocs, while MAP@10 rises by 5.30\%, 6.48\%, and 12.58\%.
The benefit therefore appears in both top-ranked relevance and deeper first-stage coverage.

\begin{figure*}[t]
    \centering
    \includegraphics[width=.88\textwidth]{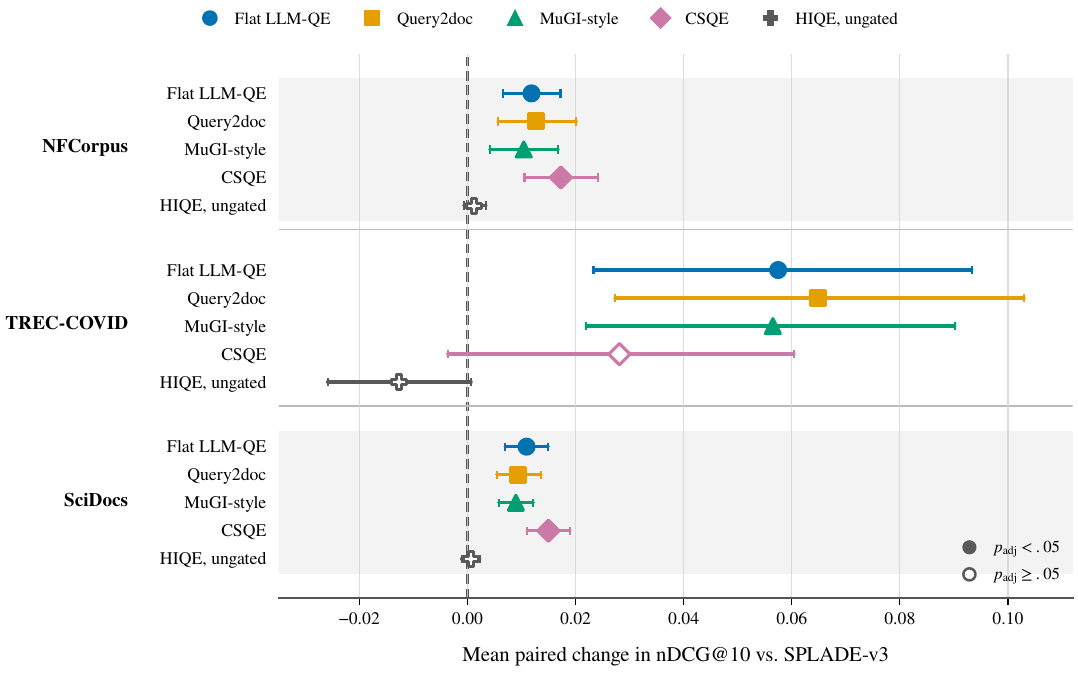}
    \caption{Paired nDCG@10 differences from SPLADE-v3. Points show mean per-query differences, and bars show paired 95\% bootstrap intervals. Filled markers indicate significance after multiple-comparison correction, while hollow markers indicate nonsignificant effects. The graph condition denotes ungated \hiqe.}
    \label{fig:paired-effects}
\end{figure*}

\begin{table*}[t]
\centering
\caption{Compact paired per-query nDCG@10 analysis comparing, for each collection, the strongest aggregate generated condition and ungated \hiqe with SPLADE-v3.}
\label{tab:paired-significance}
\small
\setlength{\tabcolsep}{5pt}
\begin{tabular}{llrrrrr}
\toprule
Dataset & Method & $\Delta$ & 95\% CI & $p_{\mathrm{adj}}$ & $d_z$ & W/T/L \\
\midrule
NFCorpus & CSQE & +0.0173 & [+0.0106, +0.0241] & \textbf{4e-05} & +0.28 & 99/175/49 \\
NFCorpus & \hiqe, ungated & +0.0013 & [-0.0007, +0.0034] & 0.661 & +0.07 & 40/246/37 \\
\midrule
TREC-COVID & Query2doc & +0.0649 & [+0.0273, +0.1030] & \textbf{0.00540} & +0.47 & 34/5/11 \\
TREC-COVID & \hiqe, ungated & -0.0126 & [-0.0259, +0.0007] & 0.207 & -0.26 & 16/12/22 \\
\midrule
SciDocs & CSQE & +0.0150 & [+0.0111, +0.0189] & \textbf{4e-05} & +0.24 & 262/611/127 \\
SciDocs & \hiqe, ungated & +0.0006 & [-0.0011, +0.0022] & 1.000 & +0.02 & 77/843/80 \\
\bottomrule
\end{tabular}
\parbox{0.97\textwidth}{\scriptsize Generated-method rows use Holm adjustment across the four generated methods within that collection. Ungated graph rows use Bonferroni adjustment across the three collection-level graph comparisons.}
\end{table*}

\paragraph{Evidence across queries.}
All twelve paired mean differences are positive, and eleven remain significant after Holm correction.
The only exception is CSQE on the 50-topic TREC-COVID collection (95\% interval $[-0.0036,0.0604]$, adjusted $p=0.0944$).
Across the twelve comparisons, paired effect sizes range from approximately $d_z=0.14$ to $0.47$.

For the best method on each collection, wins outnumber losses among non-tied queries: 99 versus 49 for NFCorpus CSQE, 34 versus 11 for TREC-COVID Query2doc, and 262 versus 127 for SciDocs CSQE (Table~\ref{tab:paired-significance}).
Many queries remain unchanged, especially on NFCorpus and SciDocs, but the aggregate gains are not driven by only a few large improvements.

\subsection{RQ2: What Makes the Gain Robust?}
\label{sec:weight-robustness}
\noindent\textbf{Answer to RQ2.}
The gain does not depend on one favorable interpolation weight or on the exact form of the generated text. It persists when the original query retains at least 30\% of the mixture weight and when the generated content is shuffled or reduced to a lexical bag. The most stable source of improvement is therefore the added vocabulary.

\begin{figure*}[t]
\centering
\includegraphics[width=.97\textwidth]{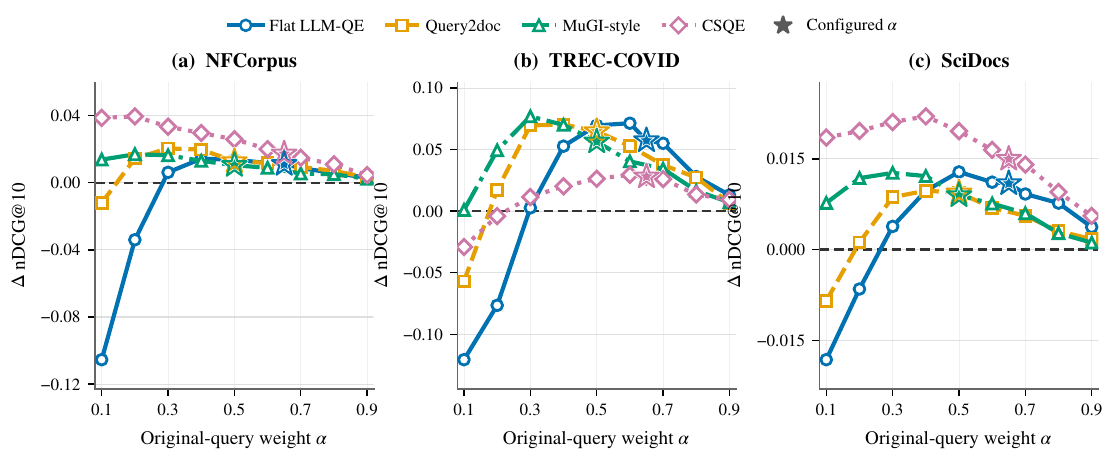}
\caption{Interpolation sensitivity versus SPLADE-v3 as the original-query weight $\alpha$ varies. Stars mark the configured weights. Most method--collection pairs remain above baseline across a broad range, while low $\alpha$ degrades Flat LLM-QE, Query2doc, and TREC-COVID CSQE.}
\label{fig:alpha-sensitivity}
\end{figure*}

\paragraph{Weight sensitivity.}
Across 114 interpolation settings, 103 beat SPLADE-v3 (Fig.~\ref{fig:alpha-sensitivity}).
MuGI-style is positive at all 27 evaluated method--collection weights, CSQE at 28/30, Flat LLM-QE at 24/30, and Query2doc at 24/27.
Every failure occurs at $\alpha=.1$ or $.2$: Flat LLM-QE fails at both values on all three collections, Query2doc at $.1$ on all three, and CSQE at $.1$ and $.2$ on TREC-COVID.
Consequently, all 90 settings with $\alpha\geq .3$ remain above baseline.
Generated content can receive substantial weight, but letting it dominate the original query is risky.

\begin{table}[t]
\centering
\caption{Interpolation robustness and same-content controls. ``Positive $\alpha$'' counts tested weights above SPLADE-v3. BOW and Shuffle are absolute $\Delta$nDCG@10 at the configured $\alpha$.}
\label{tab:robustness-controls}
\footnotesize
\setlength{\tabcolsep}{3.0pt}
\begin{tabular}{lrrr}
\toprule
Method & Positive $\alpha$ & BOW $\Delta$ & Shuffle $\Delta$ \\
\midrule
\multicolumn{4}{l}{\emph{NFCorpus}} \\
Flat LLM-QE & 8/10 & +.0094 & +.0104 \\
Query2doc & 8/9 & +.0107 & +.0141 \\
MuGI-style & 9/9 & +.0085 & +.0104 \\
CSQE & 10/10 & +.0180 & +.0167 \\
\midrule
\multicolumn{4}{l}{\emph{TREC-COVID}} \\
Flat LLM-QE & 8/10 & +.0620 & +.0604 \\
Query2doc & 8/9 & +.0358 & +.0585 \\
MuGI-style & 9/9 & +.0469 & +.0557 \\
CSQE & 8/10 & +.0219 & +.0238 \\
\midrule
\multicolumn{4}{l}{\emph{SciDocs}} \\
Flat LLM-QE & 8/10 & +.0129 & +.0107 \\
Query2doc & 8/9 & +.0084 & +.0065$^{\dagger}$ \\
MuGI-style & 9/9 & +.0099 & +.0112 \\
CSQE & 10/10 & +.0157 & +.0159 \\
\bottomrule
\end{tabular}
\vspace{1pt}
\parbox{0.97\columnwidth}{\scriptsize $^{\dagger}$Only shuffled Query2doc on SciDocs differs significantly from its contextual counterpart after within-collection Holm correction ($p=.0434$). It remains above SPLADE-v3.}
\end{table}

\paragraph{Representation controls.}
Both same-content controls remain above SPLADE-v3 in all 24 aggregate method--collection comparisons.
Moreover, 23 of 24 controls do not differ significantly from their contextual counterpart after Holm correction.
The only exception is shuffled Query2doc on SciDocs ($\Delta=-.0029$, $p_{\mathrm{Holm}}=.0434$), which still beats the baseline.
Nine alternative representations even score higher in aggregate than the corresponding contextual representation, although none of those increases is significant after correction.
Shuffling therefore removes coherent word order, and the lexical-bag control removes contextual SPLADE expansion, yet both retain the positive effect.
The common ingredient is the vocabulary supplied by the generator.

\subsection{RQ3: Can Structured Expansion Match the Gain?}
\noindent\textbf{Answer to RQ3.}
No. Ungated \hiqe changes nDCG@10 by only $+.0013$ on NFCorpus, $-.0126$ on TREC-COVID, and $+.0006$ on SciDocs.
All three paired 95\% intervals include zero, and none is significant after Bonferroni correction.
The ablations in Table~\ref{tab:ablations} likewise reveal no consistently useful relation, weighting, validation, or depth choice.

\begin{table*}[t]
\centering
\caption{Hierarchy ablations as absolute $\Delta$nDCG@10 relative to SPLADE-v3, with mean $\Delta$ defined as the arithmetic mean across the three collections.}
\label{tab:ablations}
\small
\setlength{\tabcolsep}{5pt}
\begin{tabular}{lrrrr}
\toprule
Variant & NFCorpus & TREC-COVID & SciDocs & Mean $\Delta$ \\
\midrule
Random count-matched & $-.0001$ & $-.0060$ & $+.0003$ & $-.0020$ \\
Flat concepts, equal scores & $-.0048$ & $-.0223$ & $-.0002$ & $-.0091$ \\
No corpus validation & $+.0006$ & $-.0256$ & $+.0006$ & $-.0081$ \\
Equal relation weights & $+.0005$ & $-.0192$ & $+.0014$ & $-.0058$ \\
Parent only & $+.0002$ & $-.0054$ & $+.0005$ & $-.0016$ \\
Child only & $+.0006$ & $-.0093$ & $+.0004$ & $-.0028$ \\
Sibling only & $+.0008$ & $-.0099$ & $+.0003$ & $-.0030$ \\
Depth 1 & $+.0015$ & $-.0129$ & $+.0010$ & $-.0035$ \\
Full graph, depth 2 & $+.0013$ & $-.0126$ & $+.0006$ & $-.0036$ \\
\bottomrule
\end{tabular}
\end{table*}

\paragraph{Ablation evidence.}
Flat equal-weight concepts are worse than the full graph on all three collections.
Removing corpus validation causes the largest TREC-COVID degradation, while parent-only traversal has the least negative mean among the named graph variants.
More structure is not consistently better: the random count-matched control has a less negative mean than the full graph, and depth two is slightly worse on average than depth one.

The stored hierarchy diagnostics show concept coverage of 100.0\% on NFCorpus, 99.71\% on TREC-COVID, and above 99.99\% on SciDocs.
The archive does not include connected-component or broader/narrower cycle counts, so we make no stronger topology claim.
The available evidence is nevertheless clear: nearly complete concept coverage does not translate into a ranking gain.

\paragraph{Gating does not rescue the graph.}
The stored rankings uniquely establish that the learned gate selects graph expansion for all 50 TREC-COVID topics. It therefore preserves the TREC-COVID graph degradation, while a per-query oracle remains $0.0232$ nDCG@10 above the learned gate there.
On NFCorpus and SciDocs, exact activation counts cannot be recovered for 14 and 13 queries because the base and graph rankings are identical and the archive lacks gate-decision logs.
Among the remaining queries, inferred learned-gate activation is 37.9\% on NFCorpus and 0\% on SciDocs.
The heuristic gate selects no graph run on the unambiguous NFCorpus or SciDocs queries and none on TREC-COVID.
Neither selector therefore offers a consistent advantage over leaving SPLADE-v3 unexpanded.

\section{Discussion}
The surprising result is that generated expansion still helps after SPLADE-v3 has already performed its own learned lexical expansion.
Earlier evidence suggests that generative expansion becomes less useful as the retriever grows stronger \cite{weller-etal-2024-generative}; here, four generation formats improve a strong frozen retriever on all three collections.
Because the document index and retrieval pipeline are fixed, the difference comes from the query-side evidence.

The controls identify that evidence more precisely.
Performance remains positive across a broad range of interpolation weights, so the result is not an artifact of one favorable coefficient.
Shuffled text retains the gain, showing that coherent generated word order is not essential.
A non-contextual lexical bag also retains the gain, showing that the contextual encoding of the generated passage is not essential either.
Across the experiments, the most stable common factor is the added scientific vocabulary: entities, methods, and specialized terms that were absent from the short query.

That vocabulary should complement, not replace, the user's wording.
Every tested configuration with $\alpha\geq .3$ beats the baseline, while all failures occur when the original query receives only 10\% or 20\% of the mixture weight.
This pattern supports integration methods that preserve substantial original-query mass and adapt the expansion weight by query, as in QuDAR \cite{kim-etal-2026-qudar}.

The lexical controls also suggest a practical design choice.
A generated passage can be used internally as a source of candidate terms; it need not be displayed to the user or treated as a factual answer.
Separating retrieval utility from user-facing generation reduces the importance of the passage's prose quality while keeping attention on the terms that affect ranking.

The graph comparison shows that adding related terms is not sufficient by itself.
Corpus support, typed relations, deeper traversal, and relation-specific weighting do not reproduce the generated gains in this pipeline.
Future structured systems may need retrieval-supervised concept selection, phrase-preserving representations, and gates trained directly for per-query ranking impact.

\subsection{Limitations}
The study covers one SPLADE-v3 configuration and three English BEIR collections.
Other retrievers, languages, domains, and interactive settings may produce different gains.
The public benchmarks may also have appeared in the generator's training data; we do not use a documented-cutoff generator, analyze corpus overlap, or include a paraphrased-query control \cite{yoon-etal-2025-hypothetical}.

The primary methods use fixed interpolation weights rather than a shared development-set tuning protocol.
The sweep demonstrates a broad positive region, but it is not a substitute for prospective tuning.
Finally, the graph is built automatically and validated with corpus evidence rather than manually labeled edges.
Its negative results may reflect graph construction, graph-guided integration, or both.

\section{Conclusion}
Generated query expansion can improve a strong learned sparse retriever even when that retriever already performs contextual lexical expansion.
Across three scientific collections, all four generated formats improve SPLADE-v3, and the gains survive wide changes in mixture weight, token order, and contextual representation.
The tested concept graph does not show the same benefit.
The clearest interpretation is that generation contributes useful scientific vocabulary---but it works as an addition to a strongly preserved original query, not as a replacement for it.

\section*{Acknowledgment}
This manuscript has been approved for unlimited release and has been assigned LA-UR-26-28002. The funding for this paper was provided by Los Alamos National Laboratory (LANL). LANL is operated by Triad National Security, LLC, for the National Nuclear Security Administration of the U.S. Department of Energy (Contract No. 89233218CNA000001).

\sloppy
\balance
\bibliographystyle{ieeetr}
\bibliography{bibliography}
\end{document}